\documentclass[11pt,a4paper]{article}
\pdfoutput=1

\usepackage{jheppub}
\usepackage{amsmath,amsfonts,amssymb,amsthm}
\usepackage{physics}

\newcommand{\nn}{\nonumber}

\allowdisplaybreaks[1]

\title{Remarks on nonperturbative perturbations}
\author{Robert J. Saskowski}
\emailAdd{robert\_saskowski@tju.edu.cn}
\affiliation{Center for Joint Quantum Studies and Department of Physics,\\School of Science, Tianjin University, Tianjin 300350, China}

\begin{document}

\abstract{We consider the linearized perturbations of near-horizon extremal Reissner-Nordstr\"om black holes in $d$-dimensional Einstein-Maxwell-Gauss-Bonnet gravity and seven-dimensional third-order Lovelock gravity. We find the solutions for the gravitational perturbations as a function of the higher-derivative coupling coefficients, which we treat nonperturbatively. Consequently, we observe a breakdown in perturbation theory for large harmonics for the six-derivative corrections.}

\maketitle


\section{Introduction}
The extremal limit of black holes is generally thought to yield a smooth solution to the Einstein equations. However, it was observed in~\cite{Welch:1995dh} that static multi-black hole solutions in five dimensions are $C^2$, rather than $C^\infty$ as we might expect, and Ref.~\cite{Candlish:2007fh} found that such solutions in dimension $d>5$ have tidal force singularities on the horizon. Such tidal force singularities of extremal black holes were also observed in AdS~\cite{Dias:2011at,Maeda:2011pk,Hickling:2015ooa,Iizuka:2022igv}, which culminated in the recent observation that almost all four-dimensional extremal AdS black holes are singular~\cite{Horowitz:2022mly}. The situation for asymptotically flat extremal black holes is indeed much better than for AdS; however, even these exhibit similar instabilities once higher-derivative corrections are taken into account~\cite{Horowitz:2023xyl,Horowitz:2024dch,Cano:2024bhh}.

Here, we focus on the extremal Reissner-Nordstr\"om black hole in $d$-dimensions, which has a near-horizon $\mathrm{AdS}_2\times S^{d-2}$ geometry. In this context, Ref.~\cite{Horowitz:2023xyl} considered linearized tensor perturbations in the near-horizon region, whose functional form is entirely determined by symmetry considerations, decomposing into harmonics of $\mathrm{AdS}_2$ and $S^{d-2}$. This reduces the equations of motion for the linearized perturbations to a system of algebraic equations. Of particular note, the perturbations scale as $r^\gamma$ for some scaling dimension $\gamma$, where $r$ is the radial coordinate of the $\mathrm{AdS}_2$. As long as $\gamma$ is a non-negative integer, such perturbations are analytic as we approach the horizon at $r=0$. However, for non-integer $\gamma$, the metric is only $C^{\lfloor\gamma\rfloor}$, and, for non-zero $\gamma<1$, the metric suffers from a tidal force singularity at the horizon. The equations of motion effectively reduce to a fourth-order polynomial equation for $\gamma$, with two of the roots being physically meaningful solutions. This was done for the case of the general four-derivative action
\begin{align}
    e^{-1}\mathcal L&=R-\frac{1}{4}F^2+c_1 R^2+c_2 (R_{\mu\nu})^2+c_3 (R_{\mu\nu\rho\sigma})^2+c_4 RF^2+c_5 R_{\mu\nu}F^{\mu\rho}F^\nu{}_\rho\nn\\
    &\qquad\,+c_6 R_{\mu\nu\rho\sigma}F^{\mu\nu}F^{\rho\sigma}+c_7 (F^2)^2+c_8 F^4\,,
\end{align}
where $R_{\mu\nu\rho\sigma}$ is the Riemann tensor, $R_{\mu\nu}$ the Ricci tensor, $R$ the Ricci scalar, and $F_{\mu\nu}$ the Maxwell field strength. Including these higher-derivative corrections led to corrected scaling dimensions, for example,
\begin{align}
    \gamma_{+-}&=\frac{\ell}{d-3}-1-d_0\frac{4(d-4)\ell(\ell+d-3)}{(d-2)(d-3)^2(2\ell-d+3)r_+^2}+\mathcal{O}(c_i^2)\,,\nn\\
    d_0&=\frac{(d-3)(d-4)^2}{4}c_1+\frac{(d-3)(2d^2-11d+16)}{4}c_2+\frac{2d^3-16d^2+45d-44}{2}c_3\nn\\
    &\qquad+(d-2)(d-3)(d-4)c_4+(d-2)(d-3)^2(c_5+c_6)+2(d-2)^2(d-3)(2c_7+c_8)\,.\label{eq:pertcorr}
\end{align}
In particular, regardless of the sign of $d_0$, the higher-derivative corrections generically lead to non-integer $\gamma$, and hence non-analytic behavior on the horizon. For the particular case of $d=\ell+3$, the two-derivative $\gamma_{+-}$ is zero and the sign is fully determined by the higher-derivative corrections. For $d_0>0$, this leads to a much stronger singular behavior on the horizon.

This non-analyticity was interpreted in~\cite{Horowitz:2023xyl} as a breakdown of \textit{effective} field theory, so it is natural to ask if these instabilities persist if we treat the higher-derivative corrections nonperturbatively. Of course, we expect there to be an infinite series of higher-derivative corrections, which is computationally intractable to deal with. In general, taking a finite number of higher-derivative terms and trying to treat them nonperturbatively leads to higher-order equations of motion and, inevitably, the infamous Ostrogradsky instability. The problem lies in the fact that higher-order equations of motion require more boundary data and therefore produce additional solutions that do not reduce to the two-derivative solution in the perturbative limit. However, for certain theories, namely those that are quasitopological, the equations of motion remain second order and we can treat a finite number of higher-derivative corrections nonperturbatively. In particular, this gives us a toy model with which we may explore whether or not the instabilities are resolved. In this short note, we consider the case of Einstein-Maxwell-Gauss-Bonnet gravity in $d$-dimensions and find the nonperturbative scaling dimensions. The expressions themselves are not particularly enlightening, so we leave them to the main text and appendices. However, we find that the nonperturbative corrections are generically non-analytic and hence do not fix the instability. This suggests that the instability is a genuine breakdown of effective field theory. We also perform the calculation of nonperturbative scaling dimensions for the asymptotically AdS case for completeness, although it clearly cannot resolve the instability, as it is already present at the two-derivative level.

Since we are working nonperturbatively in $\alpha$, we can ask if there is any breakdown in the perturbative expansion. Indeed, it was observed in~\cite{Cano:2024bhh} that there is a breakdown of perturbation theory for four-dimensional Kerr black holes in the Eikonal limit. Curiously, this does not seem to occur for Reissner-Nordstr\"om black holes in the large $\ell$ limit. The two- and four-derivative terms in \eqref{eq:pertcorr} have the same scaling behavior in the $\ell\to\infty$ limit, so the four-derivative corrections are always under perturbative control. From our nonperturbative expressions, we see that this is indeed true to all orders in the $\alpha$ expansion. However, this need not always be the case. This motivates us to consider the case of extremal Reissner-Nordstr\"om black holes in third-order Lovelock gravity. There, we find a breakdown of the effective field theory expansion at the six-derivative level, where at large angular momentum the $\mathcal O(\alpha^2)$ term dominates over the $\mathcal O(\alpha)$ term. In this case, only the nonperturbative expression is valid for $\ell\gtrsim\alpha^{-1}$.

The rest of this note is organized as follows. In Section \ref{sec:flat}, we compute the perturbations for asymptotically flat extremal black holes in Einstein-Maxwell-Gauss-Bonnet gravity, and, in Section \ref{sec:AdS} we do the same for asymptotically AdS black holes. In section \ref{sec:Lovelock}, we extend our analysis to the case of asymptotically flat black holes in $d=7$ third-order Lovelock gravity. We conclude briefly in Section \ref{sec:end}. Many of the more technical expressions are relegated to the Appendices.

\section{Flat Gauss-Bonnet black holes}\label{sec:flat}
We consider $d$-dimensional Einstein-Maxwell-Gauss-Bonnet gravity without a cosmological constant
\begin{equation}
    e^{-1}\mathcal L=R-\frac{1}{4}F^2+\alpha\mathcal{X}_4\,,\qquad\mathcal{X}_4=(R_{\mu\nu\rho\sigma})^2-4(R_{\mu\nu})^2+R^2\,,
\end{equation}
where $F=\dd A$ locally. We generally expect $\alpha$ to be non-negative, which is required by unitarity for pure Einstein gravity~\cite{Cheung:2016wjt}, as well as for the known examples of heterotic~\cite{Gross:1986mw}, Type I~\cite{Tseytlin:1995bi},  and bosonic~\cite{Zwiebach:1985uq} strings. However, positivity of the four-derivative corrections to the entropy require that $\alpha\le 0$~\cite{Cheung:2018cwt}.\footnote{It should be noted that the corrections to the entropy are expected to be positive as a result of the weak gravity conjecture, but this does not necessarily always hold~\cite{Ma:2022gtm,Wu:2024iiz}.}

Due to the quasi-topological nature of the theory, it does not suffer from the infamous Ostrogradsky instability that plagues higher-derivative theories, even if $\alpha$ is treated nonperturbatively. This also has the consequence that the equations of motion are second-order in derivatives
\begin{align}
    0&=R_{\mu\nu}-\frac{1}{2}F_{\mu\rho}F_{\nu}{}^\rho+\alpha\qty(R_{\mu\rho\sigma\lambda} R_{\nu}{}^{\rho\sigma\lambda}-2R^{\rho\sigma} R_{\mu\rho\nu\sigma}-2R_{\mu\rho}R_\nu{}^{\rho}+R R_{\mu\nu})\nn\\
    &\qquad\quad-\frac{1}{2}\qty(R-\frac{1}{4}F^2+\alpha\mathcal{X}_4)g_{\mu\nu}\,,\nn\\
    0&=\nabla_\mu F^{\mu\nu}\,,\label{eq:eoms}
\end{align}
where $g_{\mu\nu}$ denotes the metric with torsion-free connection $\nabla$. The fully nonperturbative (in $\alpha$) static solution to the equations of motion was found to be given by a two-parameter family of solutions~\mbox{\cite{Wheeler:1985nh,Wiltshire:1985us}}
\begin{align}
    \dd s^2&=-f\dd t^2+\frac{\dd r^2}{f}+r^2\dd\Omega_{d-2}^2\,,\nn\\
    A&=\frac{\tilde q}{r^{d-3}}\dd t\,,\label{eq:BHsolnAnsatz}
\end{align}
where
\begin{equation}
    f=1+\frac{r^2}{2(d-3)(d-4)\alpha}\qty(1-\sqrt{1+2(d-3)(d-4)\alpha\qty(\frac{4\mu}{r^{d-1}}-\frac{(d-3)\tilde q^2}{(d-2)r^{2d-4}})})\,,\label{eq:flatffnct}
\end{equation}
and $\dd\Omega_{d-2}^2$ is the line element on $S^{d-2}$.

In the extremal limit, the near-horizon geometry is $\mathrm{AdS}_2\times S^{d-2}$, with metric
\begin{equation}
    \dd s^2=l^2\qty(-r^2\dd t^2+\frac{\dd r^2}{r^2})+r_+^2\dd\Omega_{d-2}^2\,,\label{eq:NHmetric}
\end{equation}
where $r_+$ is the horizon radius and $l$ is the $\mathrm{AdS}_2$ length. The near-horizon field strength is just a constant electric field
\begin{equation}
    F=\frac{q l}{r_+^2}\dd t\land\dd r\,,\label{eq:NHfieldStrength}
\end{equation} 
and we identify
\begin{align}
    q&=l\sqrt{2(d-2)(d-3)(r_+^2+(d-4)(d-5)\alpha)}\,,\nn\\ l&=r_+\sqrt{\frac{r_+^2+2(d-3)(d-4)\alpha}{(d-3)((d-3)r_+^2+(d-4)^2(d-5)\alpha)}}\,.\label{eq:ql}
\end{align}
This can be obtained directly since the near-horizon geometry is an exact solution of the equations of motion, \eqref{eq:eoms}. Note that in general dimensions, $r_+$ is a rather complicated function of $q$ and $\alpha$, so it is cleaner to express $q$ and $l$ implicitly in terms of $r_+$. At the two-derivative level, it is straightforward to see that the relationship is just
\begin{equation}
    r_+^2\Big\vert_{\alpha=0}=\sqrt{\frac{d-3}{2(d-2)}}q\,.
\end{equation}

Following \cite{Horowitz:2023xyl}, we consider linearized perturbations around the near-horizon geometry
\begin{equation}
    \check{g}_{\mu\nu}=g_{\mu\nu}+\delta g_{\mu\nu}\,,\qquad \check A_\mu=A_\mu+\delta A_\mu\,,
\end{equation}
where $\check g$ and $\check A$ denote the perturbed solution while $g$ and $A$ denote the original static solution \eqref{eq:BHsolnAnsatz}. We will work to linear order in static, scalar-derived perturbations. Due to the symmetry, such perturbations decompose into harmonics of $\mathrm{AdS}_2$ and $S^{d-2}$,
\begin{align}
    \delta g_{IJ}&=s Y_{\ell}\,r^\gamma g_{IJ}\,,\nn\\
    \delta g_{AB}&=r^\gamma\qty(h_L Y_\ell\, g_{AB}+h_T Y^\ell_{AB})\,,\nn\\
    \delta A&=Q Y_\ell\, r^{\gamma+1}\dd t\,,\label{eq:perturbations}
\end{align}
where $I,J$ and $A,B$ indices correspond to $\mathrm{AdS}_2$ and $S^{d-2}$, respectively, and $s$, $h_L$, $h_T$, and $Q$ are real constants. Note that $Y_\ell$ is a spherical harmonic with eigenvalue $\ell(\ell+d-3)$, and $Y^\ell_{AB}$ is the traceless part of the second derivative of $Y_\ell$,
\begin{equation}
    Y^\ell_{AB}=\frac{1}{\ell(\ell+d-3)}D_A D_B Y_\ell+\frac{1}{d-2}Y_\ell\bar g_{AB}\,,
\end{equation}
where $\bar g$ denotes the unit $S^{d-2}$ metric with (torsion-free) connection $D$. Due to the spherical symmetry, we may restrict the $Y_\ell$ to have no azimuthal dependence without loss of generality. That is, writing
\begin{equation}
    \dd\Omega_{d-2}^2=\dd\theta^2+\sin^2\theta\,\dd\Omega_{d-3}^2\,,
\end{equation}
the $Y_\ell(\theta)$ are given by Gegenbauer polynomials, 
\begin{equation}
    Y_\ell(\theta)=C_\ell^{(d-3)/2}(\cos\theta)\,.
\end{equation}
For the special cases of $d=4$ and $d=5$,\footnote{More generally, the $d=5$ spherical harmonics are known only recursively. See \cite{Lindblom:2017maa}.} these become the Legendre polynomials $P_\ell(\cos\theta)$ and the Chebyshev polynomials $U_\ell(\cos\theta)=\sin((\ell+1)\theta)/\sin\theta$, respectively. More generally, they can always be written as a sum of Legendre or Chebyshev polynomials using the recursion relation
\begin{equation}
    C_\ell^{n+1}(\cos\theta)=\frac{(2n+\ell+1)\cos\theta\, C^n_{\ell+1}(\cos\theta)-(\ell+2)\,C^n_{\ell+2}(\cos\theta)}{2n\sin\theta}\,.
\end{equation}

In general, there are several solution branches. At both the two- and four-derivative level, we must always solve the equation
\begin{equation}
    \gamma(\gamma-1)h_L=0\,.
\end{equation}
The case $\gamma=1$ is subtle as it renders the equations of motion underdetermined, so we focus on the case $h_L=0$.

As a consequence of Birkhoff's theorem, when $\ell=0$, the only solutions are perturbations towards other Reissner-Nordstr\"om black holes with infinitesimally changed mass and charge. The $\ell=1$ modes are special and must be treated independently since $Y^1_{AB}=0$, and hence all equations proportional to it are automatically satisfied. Thus, we will focus on $\ell\ge 2$.

With these two points in mind, at the two-derivative level, the scaling dimensions are given by
\begin{equation}
    \gamma_{\pm\pm}=\frac{1}{2}\qty(-1\pm \frac{\sqrt{5(d-3)^2+4\ell(\ell+d-3)\pm 4(d-3)(2\ell+d-3)}}{d-3})\,.
\end{equation}
Even though the equations of motion are only second-order in derivatives, they are coupled, which is why we have four roots for $\gamma$ rather than two as we might otherwise expect.
Although these roots na\"ively look non-integer, they have the simple expression
\begin{equation}
    \gamma_{--}=-\frac{\ell}{d-3}\,,\qquad\gamma_{-+}=-2-\frac{\ell}{d-3}\,,\qquad\gamma_{+-}=\frac{\ell}{d-3}-1\,,\qquad\gamma_{++}=\frac{\ell}{d-3}+1\,.
\end{equation}
As noted in \cite{Horowitz:2022mly}, we are always free to choose boundary conditions on the horizon to remove two of the four solutions. $\gamma_{--}$ and $\gamma_{-+}$ are always negative and hence blow up at the horizon, and therefore we will choose to remove these, in keeping with the conventions of~\cite{Horowitz:2022mly}. It should be noted that for $d>5$, $\gamma_{+-}$ is already negative at the two-derivative level when $\ell<d-3$. For such solutions, the linearized ansatz breaks down near the horizon.

It is possible to compute the four-derivative correction in general dimensions if $\alpha$ is treated perturbatively. Notably,
\begin{equation}
    \gamma_{+-}=\frac{\ell}{d-3}-1+\frac{\alpha}{r_+^2}\frac{(d-4)^2(3d-7)}{(d-3)^2}\frac{\ell(\ell+d-3)}{2\ell-d+3}+\mathcal{O}(\alpha^2)\,,\label{eq:pertScaleDimsGeneral}
\end{equation}
which agrees with \eqref{eq:pertcorr} if we set $c_1=-\tfrac{1}{4}c_2=c_3=\alpha$. However, we have only managed to obtain nonperturbative expressions dimension by dimension. Note that Gauss-Bonnet is topological\footnote{However, one can get a dynamical theory by rescaling $\alpha\to\alpha/(d-4)$ and taking the limit $d\to 4$~\cite{Glavan:2019inb,Lu:2020iav,Hennigar:2020lsl}. We do not consider this here.} in $d=4$ and trivial for $d<4$, so we restrict our attention to $d\ge 5$.

For simplicity of presentation, we will specialize to the case $d=5$. The scaling dimensions for $6\le d\le 11$ are computed in much the same way and may be found in Appendix~\ref{app:flat}. For $d=5$, there is a particularly nice relationship,
\begin{equation}
    r_+^2=\sqrt{\frac{q^2}{3}+4\alpha^2}-2\alpha\,.
\end{equation}

The equations of motion, \eqref{eq:eoms}, together with the ansatz for the perturbations, \eqref{eq:perturbations}, reduces to six algebraic equations
\begin{align}
    0&=2 h_T (\ell -1) (\ell +3) \left(4 \alpha +r_+^2\right)-24 \sqrt{3} (\gamma +1) Q r_+^3\nn\\
    &\qquad+3 s \left(4 \alpha+r_+^2\right) \left(r_+^2 (\ell  (\ell +2)+12)+4 \alpha  \ell  (\ell +2)\right)\,,\nn\\
    0&=-6 \gamma h_T-\ell  (\ell +2) \left(-2
   \gamma  h_T+6 \sqrt{3} Qr_+-3 \gamma  s \left(4 \alpha +r_+^2\right)\right)\,,\nn\\
   0&=h_T r_+^2 (12 \gamma  (\gamma
   +1)+\ell  (\ell +2))-12 \alpha  h_T \ell  (\ell +2)+6 r_+^2 s \ell  (\ell +2) \left(4 \alpha +r_+^2\right)\,,\nn\\
   0&=h_T
   \left((2 \gamma +1)^2r_+^2-12 \alpha \right)+6 r_+^2 \left(\left(\gamma ^2+\gamma +1\right) s \left(4 \alpha
   +r_+^2\right)-2 \sqrt{3} (\gamma +1) Q r_+\right)\nn\\
   0&=\sqrt{3} \gamma  \left(4 \alpha +r_+^2\right) \left(h_T (\ell -1)+6 r_+^2 s (\ell +3)\right)\,,\nn\\
   &\qquad+4 Q r_+ (\ell +3)
   \left(r_+^2 (\ell  (\ell +2)-3 \gamma  (\gamma +1))+4 \alpha  \ell  (\ell +2)\right)\,,\nn\\
   0&=\sqrt{3} \gamma  \left(4 \alpha +r_+^2\right) \left(h_T (\ell -2) (\ell -1)+10 r_+^2 s \ell  (\ell +3)\right)\nn\\
   &\qquad+2 Q r_+
   \ell  (\ell +3) \left(r_+^2 (-10 \gamma  (\gamma +1)+3 \ell  (\ell +2)+1)+4 \alpha  (3 \ell  (\ell +2)+1)\right)\,.
\end{align}
Note that, following our earlier discussion, we have already set $h_L=0$. These equations are not all independent, and we find the solution
\begin{align}
    h_T&=-\frac{3 s \left(4 \alpha +r_+^2\right) \left(r_+^2 \left(7 \ell  (\ell +2)-12 \left(\gamma ^2+\gamma -2\right)\right)+12 \alpha 
   \ell  (\ell +2)\right)}{8 r_+^2 (\ell -1) (\ell +3)}\,,\nn\\
    Q&=-\frac{\sqrt{3} s \left(4 \alpha +r_+^2\right)}{32 (\gamma +1)
   r_+^5}\Big[r_+^4 \left(\ell  (\ell +2)-4 \left(\gamma ^2+\gamma +2\right)\right)+8
   \alpha  r_+^2 \left(\ell  (\ell +2)-2 \left(\gamma ^2+\gamma -2\right)\right)\nn\\
   &\kern9em+16 \alpha ^2 \ell  (\ell +2)\Big]\,,
\end{align}
leaving us to solve the equation
\begin{equation}
    \left(r_+^2 (-2 \gamma +\ell +2) (2 \gamma +\ell +4)+4 \alpha  \ell  (\ell +2)\right) \left(r_+^2 (2 \gamma +\ell ) (\ell -2
   (\gamma +1))+4 \alpha  \ell  (\ell +2)\right)=0\,.
\end{equation}
The four solutions for the scaling dimensions are thus
\begin{align}
    \gamma_{\pm\pm}=\frac{1}{2}\qty(-1\pm \frac{\sqrt{(\ell+1\pm 2)^2r_+^2+4\ell(\ell+2)\alpha }}{r_+})\,.\label{eq:5dFlatScaleDim}
\end{align}
Note that, expanded perturbatively in $\alpha$, this matches the perturbative calculation~\eqref{eq:pertScaleDimsGeneral} with $d=5$.

For example, we see that the $\ell=2$ mode has a scaling dimension
\begin{equation}
    \gamma_{+-}=\frac{1}{2}\qty(-1+\frac{\sqrt{r_+^2+24\alpha}}{r_+})<1\,,
\end{equation}
which generically results in tidal force singularities on the horizon. For large enough $\ell$, there is no tidal force singularity but the higher-derivative corrections still do not resolve the non-analyticity. Thus, the instability persists nonperturbatively in $\alpha$.

It is curious to observe that the $\ell=2$ scaling dimension is negative when $\alpha<0$, although this does not seem to be a physical choice of sign. Moreover, when $\gamma<0$, the metric blows up at the horizon, and we should not trust the linearized expansion. In the extreme case that $\alpha$ becomes negative and large enough, the scaling dimension becomes complex. Additionally, since we are working nonperturbatively in $\alpha$, we are free to take $\alpha$ large. In particular, $\alpha=n(n+1)r_+^2/6$ corresponds to $\gamma_{+-}=n$ (for $\ell=2$ and $n\ge 0$), so we can technically resolve the instability for large $\alpha$. Of course, this only resolves one instability since $\alpha$ can't depend on $\ell$.

Looking ahead, it is also interesting to note that the coefficient of $r_+^2$ and the coefficient of $\alpha$ in the square root of \eqref{eq:5dFlatScaleDim} both scale as $\ell^2$, which means that there is no breakdown in the perturbative expansion at large $\ell$ at any order in $\alpha$.

\section{AdS Gauss-Bonnet black holes}\label{sec:AdS}
For completeness, we can turn on a cosmological constant,
\begin{equation}
    e^{-1}\mathcal L=R+\frac{(d-1)(d-2)}{L^2}-\frac{1}{4}F^2+\alpha\mathcal{X}_4\,,
\end{equation}
where $L$ is the AdS radius. The full nonperturbative solution to the equations of motion still takes the same form as the flat case \eqref{eq:BHsolnAnsatz} but with a different $f$ function appearing in the metric~\cite{Wheeler:1985nh,Wiltshire:1985us}
\begin{equation}
    f=1+\frac{r^2}{2(d-3)(d-4)\alpha}\qty(1+\sqrt{1+2(d-3)(d-4)\alpha\qty(\frac{4\mu}{r^{d-1}}-\frac{(d-3)\tilde q^2}{(d-2)r^{2d-4}}-\frac{1}{L^2})})\,.
\end{equation}
Note that the function $f$ is structurally different from the flat case \eqref{eq:flatffnct}, and the two are not equivalent in the $L\to\infty$ limit. The unperturbed extremal near-horizon solution is given by an $\mathrm{AdS}_2\times S^{d-2}$ geometry \eqref{eq:NHmetric} with constant electric field \eqref{eq:NHfieldStrength} subject to the identifications
\begin{align}
    q&=l\frac{\sqrt{2(d-2)((d-1)r_+^4+(d-3)L^2(r_+^2+(d-4)(d-5)\alpha))}}{L}\,,\nn\\ l&=Lr_+\sqrt{\frac{r_+^2+2(d-3)(d-4)\alpha}{(d-1)(d-2)r_+^4+(d-3)L^2((d-3)r_+^2+(d-4)^2(d-5)\alpha)}}\,.
\end{align}
In the limit that $L\to\infty$, these reduce to the asymptotically flat case~\eqref{eq:ql}. Note that there are now two AdS radii: $l$ is the near-horizon $\mathrm{AdS}_2$ radius, whereas $L$ is the $\mathrm{AdS}_d$ radius.

At the two-derivative level, the scaling dimensions are given by
\begin{align}
    \gamma_{\pm\pm}&=\frac{1}{2}\qty(-1\pm\sqrt{\frac{5(d-1)(d-2)r_+^2+L^2(5(d-3)^2+4\ell(\ell+d-3))\pm\sqrt{b}}{(d-3)^2L^2+(d-1)(d-2)r_+^2}})\,,\nn\\
    b&=(d^2-3d+2)^2 r_+^4+(d-3)^2 L^4(2\ell+d-3)^2\nn\\
    &\quad+2(d-3) (d-1) L^2 r_+^2 [(d-3) (d-2)+2\ell(\ell+d-3)]\,.
\end{align}
If we define
\begin{equation}
    \sigma=(d-3)^2+(d-1)(d-2)\frac{r_+^2}{L^2}\,,
\end{equation}
then the expression simplifies to
\begin{align}
    \gamma_{\pm\pm}&=\frac{1}{2}\qty(-1\pm\sqrt{\frac{5\sigma+4\ell(\ell+d-3)\pm \sqrt{4\tfrac{d-3}{d-2}\ell(\ell+d-3)(\sigma+d-3)+\sigma^2}}{\sigma}})\,.
\end{align}
This is already non-analytic (and, for some values of $\ell$ and $\sigma$, singular) before the addition of any higher-derivative corrections~\cite{Horowitz:2022mly}, which is why we treat this case separately from the asymptotically flat case, even though the latter is realized as the $L\to\infty$ limit.

We present the nonperturbative higher-derivative corrections to the scaling dimensions for completeness, although they do not resolve this instability. For example, in five dimensions, the scaling dimensions are given by
\begin{align}
    \gamma_{\pm\pm}&=\frac{1}{2}\qty(-1\pm\sqrt{5+\frac{\ell(\ell+2)(1-4\chi)}{\sigma}+\frac{48\ell(\ell+2)\chi^2}{\sigma-4}\pm\frac{4\sqrt{b}}{\sigma}})\,,\nn\\
    b&=\tfrac{8}{3}\ell(\ell+2)(2 + \sigma - 6 (4 + \sigma) \chi+ 24\ell (\ell+2) \chi^2)+\sigma^2\,,
\end{align}
where we have defined $\chi=\alpha/L^2$. The scaling dimensions for higher dimensions are considerably messier and are relegated to Appendix \ref{app:AdS}.

\section{Lovelock black holes}\label{sec:Lovelock}
One can also consider Einstein-Maxwell-Lovelock \cite{Lovelock:1970zsf,Lovelock:1971yv} theories more generally, with action
\begin{equation}
    e^{-1}\mathcal L=\sum_{n=0}^{N}\alpha_n\mathcal{X}_{2n}-\frac{1}{4}F^2\,,\qquad\mathcal{X}_{2n}=\frac{1}{2^n}\delta^{\mu_1\nu_1\cdots\mu_n\nu_n}_{\rho_1\sigma_1\cdots\rho_n\sigma_n}R_{\mu_1\nu_1}{}^{\rho_1\sigma_1}\cdots R_{\mu_n\nu_n}{}^{\rho_n\sigma_n}\,,
\end{equation}
where $\mathcal{X}_{2n}$ is the Euler density in $2n$ dimensions and we conventionally normalize $\alpha_0=-2\Lambda$ and $\alpha_1=1$. Once again, due to the quasi-topological nature of the theory, we are permitted to treat the $\alpha_i$ nonperturbatively. The Einstein equation is given by
\begin{align}
    \sum_{n=0}^N\alpha_nG^{(n)}_{\mu\nu}&=\frac{1}{2}\qty(F_{\mu\rho}F_{\nu}{}^\rho-\frac{1}{4}F^2g_{\mu\nu})\,,
\end{align}
where
\begin{equation}
    G^{(n)}_{\mu\nu}=-\frac{1}{2^{n+1}}g_{\rho\mu}\delta^{\rho\alpha_1\beta_1\cdots\alpha_n\beta_n}_{\nu\gamma_1\delta_1\cdots\gamma_n\delta_n}R_{\alpha_1\beta_1}{}^{\gamma_n\delta_n}\cdots R_{\alpha_n\beta_n}{}^{\gamma_n\delta_n}\,,
\end{equation}
while the Maxwell equation remains unchanged. Exact nonperturbative black hole solutions are known \cite{Wheeler:1985qd,Zegers:2005vx} and take the usual form
\begin{align}
    \dd s^2&=-f\dd t^2+\frac{\dd r^2}{f}+r^2\dd\Omega_{d-2}^2\,,\nn\\
    A&=\frac{\tilde q}{r^{d-3}}\dd t\,,
\end{align}
where now $f$ is a root of the equation
\begin{equation}
    \sum_{n=0}^N\tilde \alpha_n\qty(\frac{1-f}{r^2})^n=\frac{4\mu}{r^{d-1}}-\frac{(d-3)\tilde q^2}{(d-2)r^{2d-4}}\,,
\end{equation}
where the different branches correspond to different cosmological constants and $\tilde\alpha_n$ are the rescaled couplings
\begin{equation}
    \tilde \alpha_0=\frac{\alpha_0}{(d-1)(d-2)}\,,\qquad\tilde \alpha_1=\alpha_1\,,\qquad\tilde\alpha_n=\alpha_n\prod_{k=3}^{2n}(d-n)\,.
\end{equation}
This is the case for spherical horizons, which we focus on here, but there are also generalizations to other horizon topologies, see \textit{e.g.} \cite{Aros:2000ij,Cai:2003kt}.

Here we focus on the case of third-order Lovelock gravity ($N=3$) in asymptotically flat space ($\alpha_0=0$), and we specialize to the case $d=7$, which is the lowest dimension for which $\mathcal{X}_6$ is non-trivial. Following our discussion from Section~\ref{sec:flat}, we might expect the physical choice of sign to be $\alpha_1>0$,\footnote{The unitarity argument~\cite{Cheung:2016wjt} can be argued perturbatively in $\alpha_i$, although the presence of a nonperturbative $\alpha_2$ could potentially complicate this argument. Nevertheless, string theoretic examples~\cite{Gross:1986mw,Tseytlin:1995bi,Zwiebach:1985uq} still suggest $\alpha_1>0$.} but, to our knowledge, the positivity bounds on $\alpha_2$ are not known. Explicitly,
\begin{align}
    \mathcal{X}_6&=2 R^{\mu \nu \sigma \kappa} R_{\sigma \kappa \rho \tau} R^{\rho \tau}{ }_{\mu \nu}+8 R^{\mu \nu}{ }_{\sigma \rho} R^{\sigma \kappa}{ }_{\nu \tau} R^{\rho \tau}{ }_{\mu \kappa}+24 R^{\mu \nu \sigma \kappa} R_{\sigma \kappa \nu \rho} R^\rho{ }_\mu+3 R R^{\mu \nu \sigma \kappa} R_{\sigma \kappa \mu \nu}\nn\\
    &\quad+24 R^{\mu \nu \sigma \kappa} R_{\sigma \mu} R_{\kappa \nu}+16 R^{\mu \nu} R_{\nu \sigma} R^\sigma{ }_\mu-12 R R^{\mu \nu} R_{\mu \nu}+R^3\,,
\end{align}
and
\begin{align}
    G^{(1)}_{\mu\nu}&=R_{\mu\nu}-\frac{1}{2}R\,g_{\mu \nu}\,,\nn\\
    G^{(2)}_{\mu\nu}&=R_{\mu\rho\sigma\lambda} R_{\nu}{}^{\rho\sigma\lambda}-2R^{\rho\sigma} R_{\mu\rho\nu\sigma}-2R_{\mu\rho}R_\nu{}^{\rho}+R R_{\mu\nu}-\frac{1}{2} \mathcal{X}_{4}\,g_{\mu \nu}\,,\nn\\
    G^{(3)}_{\mu\nu}&=-3\left(4 R^{\tau \rho \sigma \kappa} R_{\sigma \kappa \lambda \rho} R^\lambda{ }_{\nu \tau \mu}-8 R^{\tau \rho}{ }_{\lambda \sigma} R^{\sigma \kappa}{ }_{\tau \mu} R_{\nu \rho \kappa}^\lambda+2 R_\nu^{\tau \sigma \kappa} R_{\sigma \kappa \lambda \rho} R^{\lambda \rho}{ }_{\tau \mu}-R^{\tau \rho \sigma \kappa} R_{\sigma \kappa \tau \rho} R_{\nu \mu}\right.\nn\\
    &\kern3em+8 R_{\nu \sigma \rho}^\tau R^{\sigma \kappa}{ }_{\tau \mu} R_{\kappa \kappa}^\rho+8 R_{\nu \tau \kappa}^\sigma R^{\tau \rho}{ }_{\sigma \mu} R_\rho^\kappa+4 R_\nu^{\tau \sigma \kappa} R_{\sigma \kappa \mu \rho} R_\tau^\rho-4 R_\nu^{\tau \sigma \kappa} R_{\sigma \kappa \tau \rho} R^\rho{ }_\mu\nn\\
    &\kern3em+4 R^{\tau \rho \sigma \kappa} R_{\sigma \kappa \tau \mu} R_{\nu \rho}+2 R R_\nu^{\kappa \tau \rho} R_{\tau \rho \kappa \mu}+8 R_{\nu \mu \rho}^\tau R_\sigma^\rho R_\tau^\rho{ }_\tau-8 R^\sigma{ }_{\nu \tau \rho} R_\sigma^\tau R^\rho{ }_\mu\nn\\
    &\kern3em-8 R^{\tau \rho}{ }_{\sigma \mu} R_\tau^\sigma{ }_\tau R_{\nu \rho}-4 R R^\tau{ }_{\nu \mu \rho} R^\rho{ }_\tau+4 R^{\tau \rho} R_{\rho \tau} R_{\nu \mu}-8 R_\nu^\tau R_{\tau \rho} R^\rho{ }_\mu+4 R R_{\nu \rho} R_\mu^\rho{ }_\mu\nn\\
    &\kern3em-R^2 R_{\nu \mu}\Big)-\frac{1}{2} \mathcal{X}_{6}\,g_{\mu \nu}\,.
\end{align}
The near-horizon geometry is again $\mathrm{AdS}_2\times S^5$, with
\begin{equation}
    q=\frac{2\sqrt{10l^6(r_+^2+6\alpha_2)-72r_+^4\alpha_3}}{l^2}\,,
\end{equation}
and $l$ being the one positive, real root of
\begin{equation}
    72 r_+^6 \alpha_3 - 8 l^6 (2 r_+^4 + 9 r_+^2 \alpha_2 + 288 \alpha_3) + 
 l^4 (r_+^6 + 24 r_+^4 \alpha_2 + 72 r_+^2 \alpha_3)=0\,.
\end{equation}
The expression for $l$ is rather complicated, so we do not present it explicitly.

Following the same procedure as before, we find scaling dimensions
\begin{equation}
    \gamma_{\pm\pm}=\frac{1}{2}\qty(-1\pm\sqrt{\frac{\sum_{m,n=0}^2a_{mn}\alpha_2^m\alpha_3^n\pm 4\sqrt{\sum_{m,n=0}^4 b_{mn}\alpha_2^m\alpha_3^n}}{5r_+^2c}})\,,\label{eq:LovelockScaleDim}
\end{equation}
where, for example,
\begin{align}
     a_{00}&=5 l^8 r_+^8 \left(4 l^2 \ell  (\ell +4)+5 r_+^2\right)\,,\nn\\
    a_{01}&=24 l^4 r_+^4 \left(-704 l^6 \ell  (\ell +4)-5 l^4 r_+^2 (8 \ell  (\ell +4)+195)+2 l^2 r_+^4 (7 \ell  (\ell +4)+200)+175 r_+^6\right),\nn\\
    a_{10}&=12 l^8 r_+^6 \left(48 l^2 \ell  (\ell +4)+r_+^2 \left(2 \ell ^2+8 \ell +75\right)\right),\nn\\
    a_{20}&=288 l^8 r_+^4 \left(16 l^2 \ell  (\ell +4)+r_+^2 (2 \ell  (\ell +4)+25)\right),\nn\\
    b_{0,0}&=25 l^{16} r_+^{16} \left(64 l^4 \ell  (\ell +4)+r_+^4\right),\nn\\
    b_{0,1}&=120 l^{12} r_+^{12} \left[8 l^8 \ell  (\ell +4) (\ell  (\ell +4)-760)-2 l^6 r_+^2 \ell  (\ell +4) (28 \ell  (\ell +4)-3259)\right.\nn\\
    &\qquad\qquad\qquad\left.-6 l^4 r_+^4 (\ell 
   (\ell +4) (4 \ell  (\ell +4)-385)+65)+l^2 r_+^6 (160-41 \ell  (\ell +4))+70 r_+^8\right],\nn\\
    b_{1,0}&=60 l^{16} r_+^{14} \left(1360 l^4 \ell  (\ell +4)-46 l^2 r_+^2 \ell  (\ell +4)+r_+^4 (\ell  (\ell +4)+30)\right),\nn\\
    b_{2,0}&=36 l^{16} r_+^{12} \big[4 l^4 \ell  (\ell +4) (9 \ell  (\ell +4)+9800)-12 l^2 r_+^2 \ell  (\ell +4) (\ell  (\ell +4)+270)\nn\\
    &\qquad+r_+^4 (\ell  (\ell
   +4) (\ell  (\ell +4)+100)+1300)\big],\nn\\
    c&=\left(l^4 \left(-648 \alpha _3+24 \alpha _2 r_+^2+r_+^4\right)+480 \alpha _3 l^2 r_+^2+192 \alpha _3 r_+^4\right) \nn\\
    &\qquad\times\left(l^4 r_+^2 \left(12
   \alpha _2+r_+^2\right)-24 \alpha _3 \left(12 l^4+4 l^2 r_+^2+r_+^4\right)\right)\,.
\end{align}
The full list of $a$ and $b$ coefficients is rather long and unenlightening, so we have relegated it to Appendix \ref{app:lovelock}. It should be noted that, in the limit that $\alpha_2\to 0$, this reduces to the Gauss-Bonnet result~\eqref{eq:7dGBnonpertFlat}.

Interestingly, if we expand this expression perturbatively in $\alpha_2$ and $\alpha_3$ (implicitly assuming $\alpha_3$ is the same order as $\alpha_2^2$), we get
\begin{align}
    \gamma_{+-}&=\frac{\ell}{4}-1+\frac{63}{16}\frac{\ell(\ell+4)}{\ell-2}\frac{\alpha_2}{r_+^2}+\frac{3\ell(\ell+4)}{640(\ell+2)(\ell-2)^3r_+^4}\big[64 (\ell -2)^2 (\ell  (879 \ell -7804)+11676)\alpha_3\nn\\
    &\qquad-3 (\ell  (\ell  (\ell  (80 \ell +4489)+10374)-904)+28960)\alpha_2^2\big]+\mathcal{O}(\alpha^3)\,.\label{eq:pertLoveScaleDim}
\end{align}
The $\mathcal{O}(\alpha)$ term is always parametrically smaller than the $\mathcal O(\alpha^0)$ term, regardless of $\ell$. However, the $\mathcal{O}(\alpha^2)$ term dominates for large $\ell$, and hence we see that the perturbative expansion breaks down at $\ell\sim \alpha_2^{-1}$, up to $\mathcal{O}(1)$ numerical factors. A similar breakdown was observed in the Eikonal limit for Kerr black holes in \cite{Cano:2024bhh}.\footnote{Note that the Eikonal limit also involves taking the azimuthal parameter $m\to\infty$ with $\ell/m$ fixed, and so differs from the limit we take. In any case, we cannot directly compare to four dimensions since all the higher-derivative Lovelock invariants become trivial.} It should be noted that \eqref{eq:pertLoveScaleDim} is not valid for $\ell=2$, which, if treated carefully, has the expansion
\begin{align}
    \gamma_{+-}=-\frac{1}{2}+\frac{3}{2r_+}\sqrt{\frac{21\alpha_2}{2}}-\sqrt{42}\qty(1197+832\frac{\alpha_3}{\alpha_2^2})\frac{\alpha_2^{3/2}}{r_+^3}+\mathcal O(\alpha^3).
\end{align}
Nevertheless, $\gamma_{+-}$ is negative at the two-derivative level for $\ell=2$, so the linearized perturbation result should not be trusted.

Since we have the nonperturbative expression for $\gamma$, we can take the large $\ell$ limit of Eq.~\eqref{eq:LovelockScaleDim} first and then expand perturbatively in $\alpha_2$ and $\alpha_3$ to find
\begin{align}
    \gamma_{+-}\overset{\ell\to\infty}{\sim}\frac{\ell}{4}+(\ell+2)\frac{315\alpha_2-24\sqrt{25\alpha_2^2-5860\alpha_3}}{80r_+^2}+\mathcal{O}(\ell^{-1},\alpha^2)\,,
\end{align}
which involves a peculiar square root of $\alpha_3$ and a different $\alpha_3\to 0$ limit than \eqref{eq:pertLoveScaleDim}. This expression also becomes complex if $\alpha_3$ is of the same order as $\alpha_2^2$, unless $\alpha_3$ is negative. Of course, all we have done is exchange the order of limits, so this is just another manifestation of the breakdown of the perturbative expansion.

\section{Discussion}\label{sec:end}
We have computed the nonperturbative (in higher-derivative couplings) scaling dimensions of the (linearized) perturbations of near-horizon Reissner-Nordstr\"om black holes in Einstein-Maxwell-Gauss-Bonnet gravity and third-order Lovelock gravity. In particular, the horizon instabilities of \cite{Horowitz:2022mly,Horowitz:2023xyl,Horowitz:2024dch} remain in this analysis. This does not necessarily rule out the possibility of there being some particular choice of (an infinite series of) higher-derivative corrections that manages to make the instability disappear, but it does seem to suggest that such resolution would have to be fine-tuned rather than generically arising from a nonperturbative treatment of the corrections. However, to our knowledge, this is the first computation of nonperturbative (in higher-derivative couplings) scaling dimensions for black hole perturbations, which is interesting in its own right.

The main novel insight is obtained from our analysis of third-order Lovelock gravity. There, we found that the perturbative expansion breaks down for large $\ell$. In some sense, this is to be expected as high spins are associated with high energies. However, the surprising aspect is that this breakdown does not appear when only considering the four-derivative Gauss-Bonnet extended action but rather requires six-derivative Lovelock terms to become apparent. It should also be noted that the similar breakdown observed for four-dimensional Kerr black holes also involved the presence of six-derivative corrections~\cite{Cano:2024bhh}. It would be interesting to understand why such a breakdown only occurs at the six- but not four-derivative level, as well as if higher-derivative corrections further dominate over the six-derivative corrections in the large $\ell$ limit. It could also be interesting (although computationally quite challenging) to treat the perturbations at the fully non-linear level, in addition to nonperturbatively in the higher-derivative coupling.

\section*{Acknowledgements}
I would like to thank Jim Liu for many insightful discussions and for his comments on the draft. This work was supported in part by the U.S. Department of Energy under grant DE-SC0007859, in part by a Leinweber Summer Research Award, and in part by National Key Research and Development Program No. 2022YFE0134300. Part of this work was performed while I was a graduate student affiliated with the Leinweber Center for Theoretical Physics at the University of Michigan.

\appendix
\section{Asymptotically flat scaling dimensions}\label{app:flat}
Here we summarize the nonperturbative (in $\alpha$) scaling dimensions of perturbations for $6\le d\le 11$ for asymptotically flat extremal near-horizon Reissner-Nordstr\"om black holes in Einstein-Maxwell-Gauss-Bonnet gravity. These generically take the form
\begin{equation}
    \gamma_{\pm\pm}=\frac{1}{2}\qty(-1\pm\sqrt{\frac{a\pm4\sqrt{b}}{(r_+^2+2(d-4)(d-5)\alpha)(c_1r_+^2+c_2\alpha)}})\,,
\end{equation}
with dimension-dependent coefficients as follows.

\subsection*{\underline{$d=6$:}}
\begin{align}
  a&=r_+^4 (4 \ell  (\ell +3)+45)+80 \alpha  r_+^2 (\ell  (\ell +3)+3)+48 \alpha ^2 (2 \ell +1) (2 \ell +5)\,,\nn\\
   b&=9 r_+^8 (2 \ell +3)^2+96 \alpha  r_+^6 (2 \ell +3)^2+16 \alpha ^2 r_+^4 (\ell  (\ell +3) (\ell  (\ell +3)+114)+198)\nn\\
   &\qquad+4608
   \alpha ^3 r_+^2 (\ell  (\ell +3)+1)+2304 \alpha^4 (2 \ell  (\ell +3)+1)\,,\nn\\
   c_1&=9\,,\qquad c_2=12\,.
\end{align}

\subsection*{\underline{$d=7:$}}
\begin{align}
  a&=5 r_+^4 (\ell  (\ell +4)+20)+30 \alpha r_+^2 (8 \ell  (\ell +4)+55)+72 \alpha ^2 (22 \ell  (\ell +4)+75)\,,\nn\\
   b&=100 r_+^8 (\ell +2)^2+3300 \alpha  r_+^6 (\ell +2)^2+45 \alpha ^2 r_+^4 (\ell  (\ell +4) (5 \ell  (\ell +4)+1012)+3380)\nn\\
   &\qquad+1080
   \alpha ^3 r_+^2 (\ell  (\ell +4) (\ell  (\ell +4)+292)+660)+1296 \alpha ^4 (\ell  (\ell +4) (\ell  (\ell +4)+660)+900)\,,\nn\\
   c_1&=20\,,\qquad c_2=90\,.\label{eq:7dGBnonpertFlat}
\end{align}

\subsection*{\underline{$d=8:$}}
\begin{align}
   a&=r_+^4 (4 \ell  (\ell +5)+125)+8 \alpha  r_+^2 (44 \ell  (\ell +5)+525)+640 \alpha ^2 (7 \ell  (\ell +5)+45)\,,\nn\\
   b&=7680 \alpha ^3 r_+^2 (\ell  (\ell +5) (\ell  (\ell +5)+323)+1260)+64 \alpha ^2 r_+^4 (\ell  (\ell +5) (9 \ell  (\ell +5)+2885)+15525)\nn\\
   &\qquad+1680
   \alpha  r_+^6 (2 \ell +5)^2+25 r_+^8 (2 \ell +5)^2+25600 \alpha ^4 (\ell  (\ell +5) (\ell  (\ell +5)+504)+1296)\,,\nn\\
   c_1&=25\,,\qquad c_2=240\,.
\end{align}

\subsection*{\underline{$d=9$:}}
\begin{align}
   a&=7 r_+^4 (\ell  (\ell +6)+45)+70 \alpha  r_+^2 (14 \ell  (\ell +6)+255)+1200 \alpha ^2 (17 \ell  (\ell +6)+175)\,,\nn\\
   b&=441 r_+^8 (\ell +3)^2+49980
   \alpha  r_+^6 (\ell +3)^2+700 \alpha ^2 r_+^4 (\ell  (\ell +6) (7 \ell  (\ell +6)+3270)+25767)\nn\\
   &\quad+126000 \alpha ^3 r_+^2 (\ell  (\ell +6) (\ell  (\ell +6)+401)+2380)+90000 \alpha ^4 (\ell  (\ell +6) (9 \ell  (\ell +6)+4760)+19600)\,,\nn\\
   c_1&=63\,,\qquad c_2=1050\,.
\end{align}

\subsection*{\underline{$d=10$:}}
\begin{align}
   a&=r_+^4 (4 \ell  (\ell +7)+245)+24 \alpha  r_+^2 (34 \ell  (\ell +7)+875)+25200 \alpha ^2 (\ell  (\ell +7)+15)\,,\nn\\
   b&=49 r_+^8 (2 \ell +7)^2+8400
   \alpha  r_+^6 (2 \ell +7)^2+360 \alpha ^2 r_+^4 (\ell  (\ell +7) (10 \ell  (\ell +7)+6419)+69580)\nn\\
   &\quad+151200 \alpha ^3 r_+^2 (\ell  (\ell +7) (\ell  (\ell +7)+502)+4200)+1587600 \alpha ^4 (\ell  (\ell +7) (\ell  (\ell +7)+600)+3600)\,,\nn\\
   c_1&=49\,,\qquad c_2=1260\,.
\end{align}
\subsection*{\underline{$d=11$:}}
\begin{align}
   a&=3 r_+^4 (\ell  (\ell +8)+80)+420 \alpha  r_+^2 (2 \ell  (\ell +8)+69)+784 \alpha ^2 (46 \ell  (\ell +8)+945)\,,\nn\\
   b&=144 r_+^8 (\ell +4)^2+34776 \alpha  r_+^6 (\ell +4)^2+147 \alpha ^2 r_+^4 (\ell  (\ell +8) (27 \ell  (\ell
   +8)+22808)+325296)\nn\\
   &\qquad+49392 \alpha ^3 r_+^2 (\ell  (\ell +8) (5 \ell  (\ell +8)+3106)+34776)\nn\\
   &\qquad+153664 \alpha ^4 (\ell  (\ell +8) (25 \ell  (\ell
   +8)+17388)+142884)\,,\nn\\
   c_1&=48\,,\qquad c_2=1764\,.
\end{align}
Note that, expanded perturbatively in $\alpha$, these scaling dimensions match the perturbative calculation~\eqref{eq:pertScaleDimsGeneral}.

\section{Asymptotically AdS scaling dimensions}\label{app:AdS}
Here we summarize the nonperturbative (in $\alpha$) scaling dimensions of perturbations for $6\le d\le 11$ for asymptotically AdS extremal near-horizon Reissner-Nordstr\"om black holes in Einstein-Maxwell-Gauss-Bonnet gravity. These generically take the form
\begin{equation}
    \gamma_{\pm\pm}=\frac{1}{2}\qty(-1\pm\sqrt{\frac{a\pm 4\sqrt{b}}{\qty(\sigma-(d-3)^2+2(d-5)(d-4)(d-2)(d-1)\chi)\qty(\qty(\sigma-(d-3)^2)\sigma+c\chi)}})\,,
\end{equation}
where
\begin{equation}
    \sigma=(d-3)^2+(d-1)(d-2)\frac{r_+^2}{L^2}\,,\qquad\chi=\frac{\alpha}{L^2}\,,
\end{equation}
and the coefficients $a,b,c$ are dimension-dependent and identified below. Note that these reduce to the results for asymptotically flat space in the limit $L\to\infty$.

\subsection*{\underline{$d=6$:}}
\begin{align}
    a&=(\sigma -9)^2 (5 \sigma +4 \ell  (\ell +3))+80 (\sigma -9) (5 (\sigma +3)+(\sigma +11) \ell  (\ell +3))\chi  \nn\\
    &\qquad+19200 (2 \ell +1) (2 \ell
   +5) \chi ^2\,,\nn\\
   b&=(\sigma -9)^4 \left(\sigma ^2+3 \sigma  \ell  (\ell +3)+9 \ell  (\ell +3)\right)\nn\\
   &\qquad-40 (\sigma -9)^3   \left(\sigma ^2 (\ell -1) (\ell +4)-\sigma  (17 \ell  (\ell +3)+12)-120 \ell 
   (\ell +3)\right)\chi\nn\\
   &\qquad+400 (\sigma -9)^2  \big[48 (4 \sigma +3)+\sigma ^2 (\ell -1)^2 (\ell +4)^2-2 \sigma  \ell  (\ell +3) (5 \ell  (\ell +3)-32)\nn\\
   &\qquad+\ell  (\ell
   +3) (25 \ell  (\ell +3)+1896)\big]\chi^2-768000 (\sigma -9) ((\sigma -57)
   \ell  (\ell +3)-4 (\sigma +3)) \chi ^3\nn\\
   &\qquad+368640000  (2 \ell  (\ell +3)+1)\chi ^4\,,\nn\\
   c&=240\,.
\end{align}

\subsection*{\underline{$d=7$:}}
\begin{align}
    a&=(\sigma -16)^2 (5 \sigma +4 \ell  (\ell +4))+72 (\sigma -16)   (25 (\sigma +6)+2 (\sigma +24) \ell  (\ell +4))\chi\nn\\
    &\qquad+51840(22 \ell 
   (\ell +4)+75)\chi ^2\,,\nn\\
   b&=(\sigma -16)^4 \left(5 \sigma ^2+16 \sigma  \ell  (\ell +4)+64 \ell  (\ell +4)\right)\nn\\
   &\qquad-360 (\sigma -16)^3 \left(\sigma ^2 (\ell  (\ell +4)-10)-2 \sigma  (23 \ell  (\ell
   +4)+30)-400 \ell  (\ell +4)\right)\chi\nn\\
   &\qquad+6480 (\sigma -16)^2 \big[1200 (2 \sigma +3) +\sigma ^2 (\ell  (\ell +4)-10)^2-12 \sigma  \ell  (\ell +4) (\ell  (\ell +4)-50)\nn\\
   &\qquad+4 \ell 
   (\ell +4) (9 \ell  (\ell +4)+3940)\big]\chi ^2\nn\\
   &\qquad+9331200 (\sigma
   -16)  (300 (\sigma +6)+\ell  (\ell +4) ((\sigma -6) \ell  (\ell +4)+2920))\chi ^3\nn\\
   &\qquad+3359232000  (\ell  (\ell +4) (\ell  (\ell+4)+660)+900)\chi ^4\,,\nn\\
   c&=2160\,.
\end{align}

\subsection*{\underline{$d=8$:}}
\begin{align}
    a&=(\sigma -25)^2 (5 \sigma +4 \ell  (\ell +5))+112 (\sigma -25) (45 (\sigma +10)+2 (\sigma +41) \ell  (\ell +5))\chi\nn\\
    &\qquad+1128960 (7 \ell  (\ell +5)+45)\chi^2\,,\nn\\
   b&=\frac{1}{3}\left\{(\sigma -25)^4 \left(3 \sigma ^2+10 \sigma  \ell  (\ell +5)+50 \ell  (\ell +5)\right)\right.\nn\\
   &\qquad-336 (\sigma -25)^3 (\sigma +10) ((\sigma -97) \ell  (\ell +5)-18 \sigma )\chi\nn\\
   &\qquad+9408 (\sigma -25)^2 \big[6480 (2 \sigma +5)+\sigma ^2 (\ell  (\ell +5)-18)^2-2 \sigma  \ell  (\ell +5) (7 \ell  (\ell +5)-1116)\nn\\
   &\qquad+49 \ell  (\ell
   +5) (\ell  (\ell +5)+1440)\big]\chi^2\nn\\
   &\qquad+94832640 (\sigma -25) (648 (\sigma +10)+\ell  (\ell +5) (18 (\sigma +298)+(\sigma -7) \ell  (\ell +5)))\chi^3\nn\\
   &\qquad\left.+238978252800 (\ell  (\ell +5) (\ell  (\ell +5)+504)+1296)\chi^4\right\},\nn\\
   c&=10080\,.
\end{align}

\subsection*{\underline{$d=9$:}}
\begin{align}
    a&=(\sigma -36)^2 (35 \sigma +4 \ell  (\ell +6)-1080)+2240 (\sigma -36) (5 (7 \sigma -201)+(\sigma -22) \ell  (\ell +6))\chi\nn\\
    &\qquad+2150400 (17 \ell  (\ell +6)+175)\chi^2\,,\nn\\
   b&=\frac{1}{7}\left\{(\sigma -36)^4 \left(7 \sigma ^2+24 \sigma  \ell  (\ell +6)+144 \ell  (\ell +6)\right)\right.\nn\\
   &\qquad-1120 (\sigma -36)^3 \left(\sigma ^2 (\ell  (\ell +6)-28)-140 \sigma  (\ell  (\ell +6)+3)-1968 \ell  (\ell +6)\right)\chi\nn\\
   &\qquad+44800 (\sigma -36)^2 \big[11760 (4 \sigma +15)+\sigma ^2 (\ell  (\ell +6)-28)^2-16 \sigma  \ell  (\ell +6) (\ell  (\ell +6)-364)\nn\\
   &\qquad+32 \ell 
   (\ell +6) (2 \ell  (\ell +6)+7161)\big]\chi^3\nn\\
   &\qquad+602112000 (\sigma -36) (3920 (\sigma +15)+\ell  (\ell +6) (28 (4 \sigma +1059)+3 (\sigma -8) \ell  (\ell +6)))\chi^3\nn\\
   &\qquad\left.+2023096320000 (\ell  (\ell +6) (9 \ell  (\ell +6)+4760)+19600)\chi^4\right\},\nn\\
   c&=33600\,.
\end{align}

\subsection*{\underline{$d=10$:}}
\begin{align}
    a&=(\sigma -49)^2 (5 \sigma +4 \ell  (\ell +7))+432 (\sigma -49) (50 (\sigma +21)+(\sigma +87) \ell  (\ell +7))\chi\nn\\
    &\qquad+130636800 (\ell  (\ell +7)+15)\chi^2\,,\nn\\
   b&=\frac{1}{4}\left\{2 (\sigma -49)^4 \left(2 \sigma ^2+7 \sigma  \ell  (\ell +7)+49 \ell  (\ell +7)\right)\right.\nn\\
   &\qquad-864 (\sigma -49)^3 \left(\sigma ^2 (\ell  (\ell +7)-40)-5 \sigma  (41 \ell  (\ell +7)+168)-3556 \ell  (\ell +7)\right)\chi\nn\\
   &\qquad+46656 (\sigma -49)^2\big[33600 (4 \sigma +21)+\sigma ^2 (\ell  (\ell +7)-40)^2\nn\\
   &\qquad-6 \sigma  \ell  (\ell +7) (3 \ell  (\ell +7)-2080)+\ell  (\ell
   +7) (81 \ell  (\ell +7)+607600)\big]\chi^2\nn\\
   &\qquad+5643509760 (\sigma -49) (2400 (\sigma +21)+\ell  (\ell +7) (20 (3 \sigma +857)+(\sigma -9) \ell  (\ell +7)))\chi^3\nn\\
   &\qquad\left.+170659735142400 (\ell  (\ell +7) (\ell  (\ell +7)+600)+3600)\chi^4\right\},\nn\\
   c&=90720\,.
\end{align}

\subsection*{\underline{$d=11$:}}
\begin{align}
    a&=(\sigma -64)^2 (5 \sigma +4 \ell  (\ell +8))+280 (\sigma -64) (135 (\sigma +28)+2 (\sigma +116) \ell  (\ell +8))\chi\nn\\
    &\qquad+8467200 (46 \ell  (\ell +8)+945)\chi^2\,,\nn\\
   b&=\frac{1}{9}\left\{(\sigma -64)^4 \left(9 \sigma ^2+32 \sigma  \ell  (\ell +8)+256 \ell  (\ell +8)\right)\right.\nn\\
   &\qquad-2520 (\sigma -64)^3 \left(\sigma ^2 (\ell  (\ell +8)-54)-6 \sigma  (47 \ell  (\ell +8)+252)-5920 \ell  (\ell +8)\right)\chi\nn\\
   &\qquad+176400 (\sigma -64)^2 \big[326592 (\sigma +7)+\sigma ^2 (\ell  (\ell +8)-54)^2-4 \sigma  \ell  (\ell +8) (5 \ell  (\ell +8)-5886)\nn\\
   &\qquad+4 \ell 
   (\ell +8) (25 \ell  (\ell +8)+349704)\big]\chi^2\nn\\
   &\qquad+10668672000 (\sigma -64) (20412 (\sigma +28)+\ell  (\ell +8) (108 (4 \sigma +1297)+5 (\sigma -10) \ell  (\ell +8)))\chi^3\nn\\
   &\qquad\left.+161310320640000 (\ell  (\ell +8) (25 \ell  (\ell +8)+17388)+142884)\chi^4\right\},\nn\\
   c&=211680\,.
\end{align}

\section{Lovelock scaling dimension}\label{app:lovelock}
Here, we give the full expression for the nonperturbative (in $\alpha$) scaling dimensions of asymptotically flat extremal near-horizon Reissner-Nordstr\"om black holes in $d=7$ third-order Lovelock gravity. These are given by
\begin{equation}
    \gamma_{\pm\pm}=\frac{1}{2}\qty(-1\pm\sqrt{\frac{\sum_{m,n=0}^2a_{mn}\alpha_2^m\alpha_3^n\pm 4\sqrt{\sum_{m,n=0}^4 b_{mn}\alpha_2^m\alpha_3^n}}{5r_+^2c}})\,,
\end{equation}
where
\begin{align}
     a_{00}&=5 l^8 r_+^8 \left(4 l^2 \ell  (\ell +4)+5 r_+^2\right)\,,\nn\\
    a_{01}&=24 l^4 r_+^4 \left(-704 l^6 \ell  (\ell +4)-5 l^4 r_+^2 (8 \ell  (\ell +4)+195)+2 l^2 r_+^4 (7 \ell  (\ell +4)+200)+175 r_+^6\right),\nn\\
    a_{02}&=576 \left[-2376 l^{10} \ell  (\ell +4)-2 l^8 r_+^2 (4417 \ell  (\ell +4)-4050)-6 l^6 r_+^4 (541 \ell  (\ell +4)+550)\right.\nn\\
    &\qquad \quad \left.-l^4 r_+^6 (376 \ell  (\ell
   +4)+3725)-4 l^2 r_+^8 (6 \ell  (\ell +4)+325)-200 r_+^{10}\right],\nn\\
    a_{10}&=12 l^8 r_+^6 \left(48 l^2 \ell  (\ell +4)+r_+^2 \left(2 \ell ^2+8 \ell +75\right)\right),\nn\\
    a_{11}&=288 l^4 r_+^2 \left[-1144 l^6 \ell  (\ell +4)-3 l^4 r_+^2 (24 \ell  (\ell +4)+425)+4 l^2 r_+^4 (13 \ell  (\ell +4)+75)\right.\nn\\
    &\qquad\qquad\qquad\left.+2 r_+^6 (4 \ell  (\ell
   +4)+75)\right]\nn,\\
   a_{12}&=0,\nn\\
    a_{20}&=288 l^8 r_+^4 \left(16 l^2 \ell  (\ell +4)+r_+^2 (2 \ell  (\ell +4)+25)\right),\nn\\
    a_{21}&=a_{22}=0,\nn\\
    b_{0,0}&=25 l^{16} r_+^{16} \left(64 l^4 \ell  (\ell +4)+r_+^4\right),\nn\\
    b_{0,1}&=120 l^{12} r_+^{12} \left[8 l^8 \ell  (\ell +4) (\ell  (\ell +4)-760)-2 l^6 r_+^2 \ell  (\ell +4) (28 \ell  (\ell +4)-3259)\right.\nn\\
    &\qquad\qquad\qquad\left.-6 l^4 r_+^4 (\ell 
   (\ell +4) (4 \ell  (\ell +4)-385)+65)+l^2 r_+^6 (160-41 \ell  (\ell +4))+70 r_+^8\right],\nn\\
    b_{0,2}&=-144 l^8 r_+^8 \left[12 l^{12} \ell  (\ell +4) (1613 \ell  (\ell +4)+243200)-140 l^{10} r_+^2 \ell  (\ell +4) (99 \ell  (\ell +4)-50)\right.\nn\\
    &\qquad\qquad-4 l^8
   r_+^4 (\ell  (\ell +4) (6713 \ell  (\ell +4)+91305)+54225)\nn\\
   &\qquad-40 l^6 r_+^6 (\ell  (\ell +4) (240 \ell  (\ell +4)-1999)-3780)+7 l^4 r_+^8
   (8400-\ell  (\ell +4) (167 \ell  (\ell +4)-8520))\nn\\
   &\qquad\left.+20 l^2 r_+^{10} (299 \ell  (\ell +4)-600)-3300 r_+^{12}\right],\nn\\
    b_{0,3}&=6912 l^4 r_+^4 \big[24 l^{16} \ell  (\ell +4) (5899 \ell  (\ell +4)+864000)+4 l^{14} r_+^2 \ell  (\ell +4) (183881 \ell  (\ell +4)-395790)\nn\\
    &\qquad+10
   l^{12} r_+^4 (\ell  (\ell +4) (66731 \ell  (\ell +4)-1269881)-126360)\nn\\
   &\qquad+l^{10} r_+^6 (\ell  (\ell +4) (208157 \ell  (\ell
   +4)-5868370)+1033200)\nn\\
   &\qquad+l^8 r_+^8 (\ell  (\ell +4) (14663 \ell  (\ell +4)-1062170)+596700)\nn\\
   &\qquad-8 l^6 r_+^{10} (\ell  (\ell +4) (397 \ell  (\ell
   +4)+14230)+16000)\nn\\
   &\qquad-4 l^4 r_+^{12} (\ell  (\ell +4) (101 \ell  (\ell +4)+1030)+39075)+40 l^2 r_+^{14} (61 \ell  (\ell +4)-1230)-5600
   r_+^{16}\big]\,,\nn\\
    b_{0,4}&=82944 \big[14889744 l^{20} \ell ^2 (\ell +4)^2+24 l^{18} r_+^2 \ell  (\ell +4) (970423 \ell  (\ell +4)-1616760)\nn\\
    &\qquad+3 l^{16} r_+^4 (\ell  (\ell
   +4) (4846259 \ell  (\ell +4)+8515920)+3499200)\nn\\
   &\qquad+10 l^{14} r_+^6 (\ell  (\ell +4) (257135 \ell  (\ell +4)+1476408)-855360)\nn\\
   &\qquad-l^{12} r_+^8 (\ell 
   (\ell +4) (1293367 \ell  (\ell +4)+271980)+7912800)\nn\\
   &\qquad-4 l^{10} r_+^{10} (\ell  (\ell +4) (164506 \ell  (\ell +4)+103805)-141000)\nn\\
   &\qquad-4 l^8
   r_+^{12} (6 \ell  (\ell +4) (4031 \ell  (\ell +4)-17150)-768625)\nn\\
   &\qquad-16 l^6 r_+^{14} (\ell  (\ell +4) (218 \ell  (\ell +4)-17505)-110050)\nn\\
   &\qquad+16 l^4
   r_+^{16} (\ell  (\ell +4) (9 \ell  (\ell +4)+3340)+31800)+640 l^2 r_+^{18} (3 \ell  (\ell +4)+130)+6400 r_+^{20}\big]\,,\nn\\
    b_{1,0}&=60 l^{16} r_+^{14} \left(1360 l^4 \ell  (\ell +4)-46 l^2 r_+^2 \ell  (\ell +4)+r_+^4 (\ell  (\ell +4)+30)\right),\nn\\
    b_{1,1}&=-144 l^{12} r_+^{10} \big[56 l^8 \ell  (\ell +4) (3 \ell  (\ell +4)+1700)+4 l^6 r_+^2 \ell  (\ell +4) (313 \ell  (\ell +4)-41515)\nn\\
    &\qquad+2 l^4 r_+^4
   (\ell  (\ell +4) (361 \ell  (\ell +4)-20645)+8400)+l^2 r_+^6 (\ell  (\ell +4) (43 \ell  (\ell +4)+4990)-6000)\nn\\
   &\qquad-10 r_+^8 (7 \ell  (\ell+4)+270\big]\,,\nn\\
    b_{1,2}&=3456 l^8 r_+^6 \big[64 l^{12} \ell  (\ell +4) (83 \ell  (\ell +4)-50250)+2 l^{10} r_+^2 \ell  (\ell +4) (22443 \ell  (\ell +4)-218540)\nn\\
    &\qquad+45 l^8
   r_+^4 (\ell  (\ell +4) (575 \ell  (\ell +4)+7194)+6580)+5 l^6 r_+^6 (\ell  (\ell +4) (517 \ell  (\ell +4)+5224)-33600)\nn\\
   &\qquad-2 l^4 r_+^8 (\ell 
   (\ell +4) (167 \ell  (\ell +4)+15495)+42300)\nn\\
   &\qquad-4 l^2 r_+^{10} (\ell  (\ell +4) (11 \ell  (\ell +4)+1140)-600)-40 r_+^{12} (2 \ell  (\ell
   +4)-45)\big]\,,\nn\\
    b_{1,3}&=165888 l^4 r_+^2 \big[12 l^{16} \ell  (\ell +4) (24479 \ell  (\ell +4)+432000)+4 l^{14} r_+^2 \ell  (\ell +4) (134059 \ell  (\ell +4)+6825)\nn\\
    &\qquad+5
   l^{12} r_+^4 (\ell  (\ell +4) (60749 \ell  (\ell +4)-683629)-165240)\nn\\
    &\qquad+5 l^{10} r_+^6 (\ell  (\ell +4) (9840 \ell  (\ell +4)-266849)+106200)\nn\\
    &\qquad-3
   l^8 r_+^8 (9 \ell  (\ell +4) (131 \ell  (\ell +4)+1560)-132650)+4 l^6 r_+^{10} (\ell  (\ell +4) (149 \ell  (\ell +4)+8900)+900)\nn\\
    &\qquad+4 l^4
   r_+^{12} (\ell  (\ell +4) (157 \ell  (\ell +4)+2710)-13875)+8 l^2 r_+^{14} (\ell  (\ell +4) (7 \ell  (\ell +4)+305)-2550)\nn\\
    &\qquad-2400
   r_+^{16}\big]\,,\nn\\
    b_{1,4}&=0\,,\nn\\
    b_{2,0}&=36 l^{16} r_+^{12} \big[4 l^4 \ell  (\ell +4) (9 \ell  (\ell +4)+9800)-12 l^2 r_+^2 \ell  (\ell +4) (\ell  (\ell +4)+270)\nn\\
    &\qquad+r_+^4 (\ell  (\ell
   +4) (\ell  (\ell +4)+100)+1300)\big],\nn\\
    b_{2,1}&=-3456 l^{12} r_+^8 \big[40 l^8 \ell  (\ell +4) (2 \ell  (\ell +4)+335)+20 l^6 r_+^2 \ell  (\ell +4) (38 \ell  (\ell +4)-3483)\nn\\
    &\qquad+5 l^4 r_+^4
   (\ell  (\ell +4) (114 \ell  (\ell +4)-2237)+2310)+2 l^2 r_+^6 (\ell  (\ell +4) (31 \ell  (\ell +4)+1730)-1700)\nn\\
    &\qquad-2 r_+^8 (\ell  (\ell +4)
   (\ell  (\ell +4)+50)+800)\big]\,,\nn\\
    b_{2,2}&=82944 l^8 r_+^4 \big[l^{12} \ell  (\ell +4) (7127 \ell  (\ell +4)-621600)+3 l^{10} r_+^2 \ell  (\ell +4) (4589 \ell  (\ell +4)-27450)\nn\\
    &\qquad+3 l^8
   r_+^4 (\ell  (\ell +4) (1129 \ell  (\ell +4)+23340)+32475)\nn\\
    &\qquad+l^6 r_+^6 (-\ell  (\ell +4) (829 \ell  (\ell +4)-14360)-43800)-l^4 r_+^8 (3 \ell 
   (\ell +4) (9 \ell  (\ell +4)+2530)+26600)\nn\\
    &\qquad+20 l^2 r_+^{10} (\ell  (\ell +4) (3 \ell  (\ell +4)-89)-80)+4 r_+^{12} (\ell  (\ell +4) (\ell 
   (\ell +4)-20)+25)\big]\,,\nn\\
    b_{2,3}&=b_{2,4}=0\,,\nn\\
    b_{3,0}&=1728 l^{16} r_+^{10} \big[4 l^4 \ell  (\ell +4) (3 \ell  (\ell +4)+1450)-4 l^2 r_+^2 \ell  (\ell +4) (2 \ell  (\ell +4)+205)\nn\\
    &\qquad+r_+^4 (\ell 
   (\ell +4)+10) (\ell  (\ell +4)+30)\big]\,,\nn\\
    b_{3,1}&=-82944 l^{12} r_+^6 \big[28 l^8 \ell  (\ell +4) (\ell  (\ell +4)-25)+10 l^6 r_+^2 \ell  (\ell +4) (22 \ell  (\ell +4)-855)\nn\\
    &\qquad+l^4 r_+^4 (\ell 
   (\ell +4) (144 \ell  (\ell +4)-695)+2550)+l^2 r_+^6 (\ell  (\ell +4)+60) (11 \ell  (\ell +4)-10)\nn\\
    &\qquad-2 r_+^8 (\ell  (\ell +4) (\ell  (\ell
   +4)+15)+150)\big]\,,\nn\\
    b_{3,2}&=b_{3,3}=b_{3,4}=0\,,\nn\\
    b_{4,0}&=20736 l^{16} r_+^8 \left(4 l^4 \ell  (\ell +4) (\ell  (\ell +4)+300)-4 l^2 r_+^2 \ell  (\ell +4) (\ell  (\ell +4)+60)+r_+^4 (\ell  (\ell
   +4)+10)^2\right),\nn\\
    b_{4,1}&=b_{4,2}=b_{4,3}=b_{4,4}=0\,,\nn\\
    c&=\left(l^4 \left(-648 \alpha _3+24 \alpha _2 r_+^2+r_+^4\right)+480 \alpha _3 l^2 r_+^2+192 \alpha _3 r_+^4\right) \nn\\
    &\qquad\times\left(l^4 r_+^2 \left(12
   \alpha _2+r_+^2\right)-24 \alpha _3 \left(12 l^4+4 l^2 r_+^2+r_+^4\right)\right)\,.
\end{align}

\bibliographystyle{JHEP}
\bibliography{cite}

\end{document}